\documentclass[aps,prl,amsmath,amssymb,preprintnumbers,nofootinbib,twocolumn,showpacs,floatfix]{revtex4}
\usepackage{epsfig,graphics,color}
\usepackage{graphicx}
\usepackage{dcolumn}
\usepackage{bm}
\usepackage{amsmath}

\newcommand \kd {\delta}
\newcommand \ra {\rightarrow}
\newcommand \eq {{\,=\,}}

\newcommand \e {\epsilon}

\newcommand \x {\cdot}

\newcommand \A {\alpha}

\newcommand \lc {\langle}
\newcommand \rc {\rangle}
\newcommand \prt {\partial}
\newcommand \D {\Delta}

\newcommand \bvec{\left( \begin{array}{c} }
\newcommand \evec{\end{array} \right)}

\newcommand \bea{\begin{eqnarray} }
\newcommand \eea{\end{eqnarray} }
\newcommand \nn {\nonumber}
\newcommand {\be} {\begin{equation}}
\newcommand {\ee} {\end{equation}}

\newcommand {\mbx} {\mbox{}}

\newcommand {\ata} {& \times &}

\begin{document}

\title{Energy and momentum deposited into a QCD medium by a jet shower} 

\author{G.-Y.~Qin}
\affiliation{Department of Physics, The Ohio State University, Columbus, OH 43210, USA}

\author{A.~Majumder}
\affiliation{Department of Physics, The Ohio State University, Columbus, OH 43210, USA}

\author{H.~Song}
\affiliation{Department of Physics, The Ohio State University, Columbus, OH 43210, USA}

\author{U.~Heinz}
\affiliation{Department of Physics, The Ohio State University, Columbus, OH 43210, USA}

\date{ \today}

\begin{abstract}
Hard partons moving through a dense QCD medium lose energy by 
radiative emissions and elastic scatterings. Deposition of the radiative 
contribution into the medium requires rescattering of the radiated gluons.
We compute the total energy loss and its deposition into the medium 
self-consistently within the same formalism, assuming perturbative 
interaction between probe and medium. The same transport coefficients
that control energy loss of the hard parton determine how the energy
is deposited into the medium; this allows a parameter free calculation 
of the latter once the former have been computed or extracted from  
experimental energy loss data. We compute them for a perturbative medium 
in hard thermal loop (HTL) approximation. Assuming that the deposited 
energy-momentum is 
equilibrated after a short relaxation time, we compute the medium's 
hydrodynamical response and obtain a conical pattern that is strongly 
enhanced by showering. 
\end{abstract}


\maketitle




Jet quenching (the modification of hard jets in dense media) is 
one of the most studied discoveries at the Relativistic Heavy-Ion 
Collider (RHIC)~\cite{RHIC_Whitepapers}. It is expected to play a key 
role in the study of the quark-gluon plasma (QGP) produced in heavy-ion 
collisions at the Large Hadron Collider (LHC). Numerous experiments 
\cite{highpt} have established the suppression of hadrons with high 
transverse momenta; others indicate that the lost energy manifests 
itself as conical flow in the soft sector \cite{back_to_back}. 

Calculations of jet modification tend to focus on one of two separate 
questions: the modification of the final hadron distribution 
from the hard parton due to its {\em energy loss}, or the response of 
the medium to the {\em energy deposited}. Numerous studies of the former, 
based on perturbative QCD (pQCD), have yielded near-rigorous measures of 
the two non-perturbative transport coefficients 
$\hat{q}\eq\frac{dp_\perp^2}{dL}$ and $\hat{e}\eq\frac{dE}{dL}$ which 
codify the transverse (to the jet axis) momentum diffusion and 
longitudinal drag experienced by a fast parton \cite{Majumder:2007iu}. 
Computations of the medium response consist of two parts: an {\em
ansatz} for the space-time profile of the energy-momentum deposition, and 
a calculation of the dynamical response to this ``source'' of excess 
energy and momentum. Based on its success at RHIC, ideal fluid dynamics 
has been used to compute this medium response \cite{hydro_mach}, 
assuming that the energy lost by the jet is entirely deposited into the 
medium at a constant rate and thermalizes instantaneously.

So far there exists no first principles calculation of the magnitude and 
space-time profile of the energy-momentum {\em deposited} in a medium by 
a hard parton that can be considered on par with the pQCD energy {\em loss} 
calculations \cite{string_mach}. A noteworthy attempt to calculate the 
deposition profile in pQCD is the semi-phenomenological approach of 
Neufeld and M\"uller \cite{Neufeld:2009ep} who use the differential 
single gluon emission spectrum of Ref.~\cite{radiative} and interpret 
this as the rate of gluon emission in the medium. A non-diffusive 
Fokker-Planck equation is then motivated to compute how this distribution 
changes due to elastic energy loss of the emitted gluons. 
As anticipated in \cite{Majumder:2004pt}, they
find that not all of the energy lost to gluon radiation is deposited in
the medium. However, since the underlying formalism \cite{radiative} lacks 
information about virtuality evolution, this calculation does not include
gluon multiplication by showering, {\it i.e.} the splitting of a radiated
gluon into two lower virtuality gluons. The transverse momentum deposition    
thus cannot be computed and, due to the strict eikonal limit used in 
\cite{radiative}, the parent parton does not lose energy after radiation.
We present a new formalism in which the radiative and elastic energy loss 
of the fast parton, its virtuality evolution by radiation, the showering 
and multiplication of the radiated gluons, and the energy deposited by them 
in the medium are all calculated consistently in 
the same approach.

Hard jets in vacuum or in heavy ion collisions are produced with 
considerable virtuality. As the jets proceed through vacuum or 
medium, this virtuality is lost by sequential radiative emissions. 
The effect of this perturbative shower on the non-perturbative 
hadronization process is computed using 
DGLAP evolution equations \cite{AP} for the fragmentation function.
These equations express the radiation of multiple partons, which 
hadronize independently, via an evolution in virtuality of 
the parent parton. In a medium, one can derive analogous equations 
where the gluon radiation probability is modified by the scattering 
of hard parton and emitted gluons off medium constituents. These 
are referred to as ``medium modified evolution equations''. 
In addition to stimulating gluon emission, the scattering of the hard 
parton causes it to lose forward light-cone momentum by elastic exchanges 
with the medium \cite{elastic,Majumder:2008zg}. At the same time the 
parton gains transverse momentum from the medium \cite{Majumder:2007hx} 
and imparts to it an equal amount in return. In an arbitrary medium 
these effects are encoded in two non-perturbative transport coefficients, 
$\hat{q}$ and $\hat{e}$, defined in terms of in-medium gluon field 
correlation functions \cite{Majumder:2007hx,Majumder:2008zg}. The medium 
modification of the standard vacuum evolution depends on these transport 
coefficients. 

In-medium evolution equations where the medium modified fragmentation 
function (MMFF) is affected only by $\hat{q}$ were derived in 
\cite{Majumder:2009zu}. We point out that the same processes can be used 
to compute the amplification of the energy deposited through multiple 
radiations stimulated by transverse broadening. Formally, this can be 
computed by replacing the operator expression for the fragmentation 
function with that for the energy deposited; this is identical to $\hat{e}$. 
Using this calculation of the modified $\hat{e}$ as the energy deposited 
through all elastic scatterings of the shower places it on the same 
footing as energy loss. The diagrams involved and the resulting 
expressions for the in-medium splitting functions (IMSF) are identical. 
The solution of the in-medium evolution equation for $\hat{e}$ no 
longer represents the elastic energy loss by one parton, but rather 
the energy deposited by the jet shower.

Imagine a hard quark or gluon with large light cone momentum $q^-$ (and 
thus energy $E = q^-/\sqrt{2}$) and virtuality $\leq \mu$ entering a 
medium of fixed length $L$ held at a constant temperature $T$. Let us 
assume that the rate of energy deposition by this jet in the medium as 
a function of length $\zeta$, denoted as $\frac{d \D E}{d\zeta} 
(E,\zeta,\mu^2)$, is known (i.e., can be calculated or measured). Note 
that both the deposited energy $\D E$ and $\zeta$ are actually the 
light-cone quantities $\D q^-$ and $\zeta^-$. For brevity we refer to 
these simply as deposited energy and distance travelled. Given the above 
function, the total energy deposited by a jet originating at location 
$\zeta_i$ and propagating to $\zeta_f$ is given as 
\bea
\D E (E,\mu^2)_{\zeta_i}^{\zeta_f} \!\!
= \!\!\!\int_{\zeta_i}^{\zeta_f}\!\!\!\! d \zeta \frac{d \D E }{d \zeta } 
  (E,\zeta,\mu^2)
\overset{1\,\mathrm{parton}}{\simeq}\! 
(\zeta_f{-}\zeta_i)\, \hat{e} , 
\label{E_L_1}
\eea
where the last approximate equality is solely for the case of a single 
parton propagating without radiation.

If the scale $\mu$ is much larger than $\Lambda_{QCD}$, the change with 
virtuality in the partonic shower pattern may be calculated perturbatively: 
a leading quark at the higher virtuality may split into a quark and a 
gluon with lower virtuality, and similarly for a gluon. As a result, there 
is change in the energy deposited in the medium due to the increase of the 
number of partons depositing energy. Using the IMSF from
\cite{Majumder:2009zu}, the change in the energy deposition by a quark 
with energy $E$ from $\zeta_i$ to $\zeta_f$ due to the increase in 
virtuality $\mu$ can be expressed as \cite{tobepublished}
\bea
\mbox{}\!\!\!\!&&\!\!\!\!\!\!\!\!\!
\frac{d \D E_q (E,\mu^2)_{\zeta_i}^{\zeta_f} }{d \ln (\mu^2)} = 
\frac{\A_s (\mu^2)}{2\pi} 
\!\!\!\int_0^1 \!\!\!dy\!\! 
\int_{\zeta_i}^{\zeta_f} \!\!\!\! d \zeta P_{q \ra qg} (y,\zeta, \mu^2, E) 
\label{quark_evol} \\
\mbx\!\!\!\ata \!\!\!\!\!\left[ \D E_q (E,\mu^2 )_{\zeta_i}^{\zeta} 
\!\!+\!\! \D E_q (yE,\mu^2 )_{\zeta}^{\zeta_f} 
\!\!+\!\! \D E_g ( (1\!\!-\!\!y)E,\mu^2 )_{\zeta}^{\zeta_f}  \right]. \nn
\eea
Here the first term in square brackets represents the energy deposited by 
a quark with energy $E$ and virtuality $\mu^2$, from the initial location 
$\zeta_i$ to the intermediate location $\zeta$; the second and third terms 
represent the energy deposited by the quark and the emitted gluon with 
reduced energies $yE$ and $(1-y)E$, respectively, from the intermediate 
location $\zeta$ to the final location $\zeta_f$. In Eq.~\eqref{quark_evol} 
the quark IMSF $P_{q\ra qg}$ is given as \cite{Majumder:2009zu,HT}
\bea
P_{q \ra qg} =  \frac{\hat{q}C_F}{2\pi \mu^2 } \frac{1{+}y^2}{1{-}y} 
\left[ 2 - 2 \cos\left( \frac{\mu^2 \zeta}{2 E y(1{-}y)} \right) \right]. 
\label{P}
\eea
The increase in the energy deposited due to the splitting of the parton 
is reduced by the virtual correction which restores unitarity to 
the evolution equations. The effect of such corrections on 
Eq.~\eqref{quark_evol} is incorporated by subtracting from it the virtual 
term  
\bea
V \!\! = \frac{\A_s (\mu^2)}{2\pi} \D E_q (E,\mu^2)_{\zeta_i}^{\zeta_f}
\!\!\!\int_0^1 \!\!\!dy\!\! 
\int_{\zeta_i}^{\zeta_f} \!\!\!\! d \zeta P_{q \ra qg} (y,\zeta, \mu^2, E). 
\label{V} 
\eea
Along with the energy deposition from a quark jet one has to evolve the 
one from a gluon jet of virtuality ${\leq\,}\mu$, using a similar evolution 
equation that includes the splitting of a gluon into two gluons or a 
$q \bar{q}$ pair. Similar to the MMFFs, one solves a coupled set of  
evolution equations for $\D E_q(E,\mu^2)_{\zeta_i}^{\zeta_f}$ and 
$\D E_g (E,\mu^2)_{\zeta_i}^{\zeta_f}$ both of which are functions 
of three variables $E,\zeta_i,\zeta_f$ at the scale $\mu^2$. 

The evolution equations (\ref{quark_evol},\ref{P},\ref{V}) for a quark 
jet and the coupled equations for gluon jets are motivated by existing 
rigorous derivations of the medium modification of 
the fragmentation functions due to gluon emission~\cite{HT}, the 
accumulation of transverse momentum \cite{Majumder:2007hx} and longitudinal 
drag~\cite{Majumder:2008zg} by propagating hard partons in a QCD medium, 
and the effect of such accumulated momentum on radiative processes
\cite{Majumder:2007ne}. The IMSF \eqref{P} accounts for interference 
between diagrams where the gluon is emitted at the origin or at the 
location $\zeta$. In propagating up to $\zeta$ the quark loses a fraction 
of its energy; while this is included in the total energy deposited, its 
effect on the interference pattern in Eq.~\eqref{P} is ignored; this is 
justified in the eikonal limit for the propagating parton. Yet another 
approximation is the neglect of the energy lost by the radiated (reabsorbed) 
gluon in the virtual correction. Since the radiated gluon in the virtual 
correction exists in only one amplitude, with a single parton in the 
complex conjugate, its energy loss is balanced by the quark propagating 
in the loop. 

In the eikonal approximation, the hard jet loses light cone momentum and 
remains close to on-shell, thus the $z$-component of the deposited 
light-cone momentum is approximately equal to the energy deposited 
($\D p_z \simeq \D E$). Note that the negative light cone momentum 
($\D q^-$) is not conjugate to $\zeta^-$ and thus it is not inconsistent 
to compute the $\zeta^-$ dependence of the $\D q^-$ deposited. The 
remaining two components that may be computed are the transverse momentum 
deposited by the jet as a function of $\zeta^-$. This can again be 
directly estimated from a pQCD calculation: A parton traversing a medium 
gains transverse momentum squared with length as  
\bea
\lc p_\perp^2\rc(E,\mu^2)_{\zeta_i}^{\zeta_f} = \int_{\zeta_i}^{\zeta_f} \!\!\!\!d\zeta {\frac{d \lc p_\perp^2 \rc(E,\mu^2) }{d\zeta}}
%
\overset{1\,\mathrm{parton}}{\simeq} 
(\zeta_f{-}\zeta_i)\, \hat{q}.
\eea
By momentum conservation this equals the $p_\perp^2$ deposited in the 
medium by the same parton. 

For a hard virtual quark the total transverse momentum deposited increases 
due to parton splitting. This can be calculated using an equation similar to 
that for light-cone momentum deposition. For a quark with energy $E$ and 
virtuality $\mu^2$, traversing a medium from $\zeta_i$ to $\zeta_f$, the 
change of the transverse momentum deposited with virtuality is obtained as 
\bea
\mbox{}\!\!\!\!&&\!\!\!\!\!\!\!\!\!
\frac{d \lc p_\perp^2 \rc_q (E,\mu^2)_{\zeta_i}^{\zeta_f} }{d \ln (\mu^2)} = 
\frac{\A_s (\mu^2)}{2\pi} 
\!\!\!\int_0^1 \!\!\!dy\!\! 
\int_{\zeta_i}^{\zeta_f} \!\!\!\! d \zeta P_{q \ra qg} (y,\zeta, \mu^2, E) 
\label{qhat_evol} \\
\mbx\!\!\!\ata \!\!\!\!\!\left[ \lc p_\perp^2 \rc_q (E,\mu^2 )_{\zeta_i}^{\zeta} 
\!\!+\!\! \lc p_\perp^2 \rc_q (yE,\mu^2 )_{\zeta}^{\zeta_f} 
\!\!+\!\! \lc p_\perp^2 \rc_g ( (1\!\!-\!\!y)E,\mu^2 )_{\zeta}^{\zeta_f} \right]. 
\nn
\eea
The splitting function here is identical to that in Eq.~\eqref{P}, 
and the meaning of the three terms in the bracket is analogous to 
Eq.~\eqref{quark_evol}. Further, one must include a virtual correction 
and couple Eq.~(\ref{qhat_evol}) to a similar equation for the $p_\perp^2$ 
deposited by a virtual gluon. 

\begin{figure}[htbp]
\includegraphics[width=0.9\linewidth]{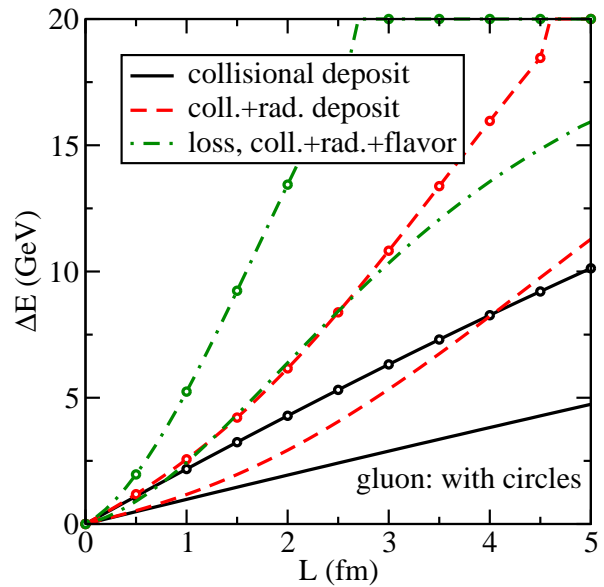} 
\caption{(Color online) Dash-dotted: Total energy lost by a hard gluon 
(circled) or quark by radiative, elastic and flavor changing processes. 
Solid: Energy deposited in the medium by a hard parton which does not 
radiate. Dashed: The same for a virtual parton devolving into a partonic 
shower.
} 
\label{fig2}
\end{figure}

\begin{figure}[htbp]
\includegraphics[width=0.9\linewidth]{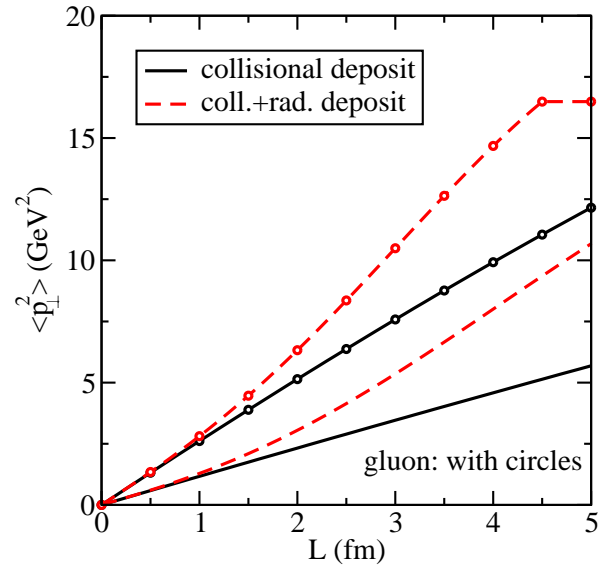}
\caption{(Color online) The $\lc p_\perp^2 \rc$ deposited by a hard gluon 
(circled) or quark without radiative emission (solid) and with a full 
radiative shower (dashed). } 
\label{fig3}
\end{figure}

Using Eqs.~(\ref{quark_evol},\ref{qhat_evol}) (along with the coupled 
ones for gluon jets), we can compute the 3-momentum $\D q^-,\, \vec{p}_T$ 
deposited by a hard virtual parton, disintegrating into a shower of 
partons, in a dense medium as a function of the length $\zeta^-$ traversed. 
Similar to the case of in-medium evolution equations for the 
MMFF~\cite{Majumder:2009zu}, these equations require an initial condition.
For the case of the MMFF, the only possible choice was to insist that the 
part of the jet with virtuality below a minimum $\mu_0^2$ exited the medium 
and use the known vacuum FF at that scale as an input. For the deposited 
part of the energy-momentum we here assume that the medium is weakly 
coupled, thus when the virtuality of the parton is $\mu_0 \simeq 4T$ 
the deposited energy and $p_\perp^2$ can be obtained from the expressions for 
$\hat{e}$ and $\hat{q}$ in an HTL plasma \cite{elastic}:  
\bea
d \D E (\mu_0, E)/d \zeta &=& C_R \A_s (\mu_0^2) m_D^2 
\log\left[  (4E T / m_D^2)^{1/4} \right], 
\nn \\
d \lc p_\perp^2 \rc (\mu_0, E)/d \zeta &=& C_R \A_s(\mu_0^2) T m_D^2 
\log\left[ 4E T /m_D^2 \right]. 
\label{input}
\eea 
Here $m_D$ is the Debye screening length and $C_R$ is the representation 
specific Casimir. The integrated energy and $p_\perp^2$ deposited from 
Eq.~\eqref{input}, as a function of the length traveled, is plotted
for gluons (circled) and quarks as solid lines in Figs.~\ref{fig2} and 
\ref{fig3}. For a consistent description, we impose that partons with an 
energy $E<4T$ become part of the thermal medium. This condition is 
maintained through the evolution equations. 

%
%
%
Using Eq.~\eqref{input} as input, we may calculate the increase in the 
energy and $p_\perp^2$ deposition in the medium as a function of $\zeta$
for initially highly virtual hard partons that evolve into a radiative 
shower. Starting from the scale of $\mu_0\eq4T$ (in all calculations we 
pick $T\eq300$ MeV and a partonic plasma with 3 quark flavors) we evolve 
up to an initial scale $\mu = E/2$. These are plotted as dashed lines 
in Figs.~\ref{fig2} and \ref{fig3}. One notes immediately that both 
quantities increase as we evolve up in virtuality. For comparison, we 
also estimate the total energy lost by the hard parton due to elastic, 
radiative inelastic and flavor changing interactions (dash-dotted lines in 
Fig.~\ref{fig2}). The last type of energy loss refers to the case where 
a quark splits with the gluon carrying a larger fraction of the momentum, 
or a gluon splits into a quark-antiquark; in this case we assume that the 
entire energy of that parent parton has been lost. This leads to a 
somewhat artificial enhancement of the total energy loss.   

As an illustration of the effect of this energy-momentum deposition in 
the medium, we compute its hydrodynamic response to the following source 
term:
\bea
J^\mu \equiv \left[ \frac{d \D E (\mu, E)}{d \zeta}, 0 , 0 , 
                    \frac{d p_z (\mu , E)}{d \zeta} \right] 
             \kd^2 (\vec{r}_\perp) \kd(t{-}z). 
\label{source}
\eea  
In this first attempt we ignore the transverse momentum contribution 
to the source current. Following Refs.~\cite{linear_hydro}, we assume that 
the energy deposited is a small perturbation and solve for the linear 
response of the medium:
\bea
T^{\mu \nu}\simeq T_0^{\mu \nu} + \kd T^{\mu \nu }; \quad  
\prt_\mu T_0^{\mu \nu } = 0 ,\quad \prt_\mu \kd T^{\mu \nu} = J^\nu. 
\label{hydro_eqn}
\eea 
$T_0^{\mu \nu}$ is the unperturbed energy-momentum tensor of a homogeneous 
and static partonic medium in equilibrium. The small excess $\kd T^{\mu \nu}$ 
is decomposed as
\bea
\kd T^{00} &\equiv& \kd \e , \quad \kd T^{0i} \equiv g^i, 
\label{t_mu_nu_decomp} \\
\kd T^{ij} &=& \kd_{ij} c_s^2 \kd \e - \Gamma_s
 \left( \prt^i g^j + \prt^j g^i - 
        \textstyle{\frac{2}{3}} \kd_{ij} \nabla \x \vec{g} \right). 
\nn
\eea
$\kd \e$ is the excess energy density, $\vec{g}$ is the momentum 
current density and $\Gamma_s\eq\frac{\eta}{sT}$ is the sound 
attenuation length. For the specific shear viscosity we took 
$\frac{\eta}{s}\eq\frac{1}{2\pi}$. We delay the response to the source 
$J$ by a time $\tau_\mathrm{rel}=\frac{1}{m_D}$ to account for 
thermalization of the deposited energy.

\begin{figure}[thbp]
\hspace{0.4cm}
\resizebox{1.1in}{1.1in}
          {\includegraphics[1in,0in][4in,3in]{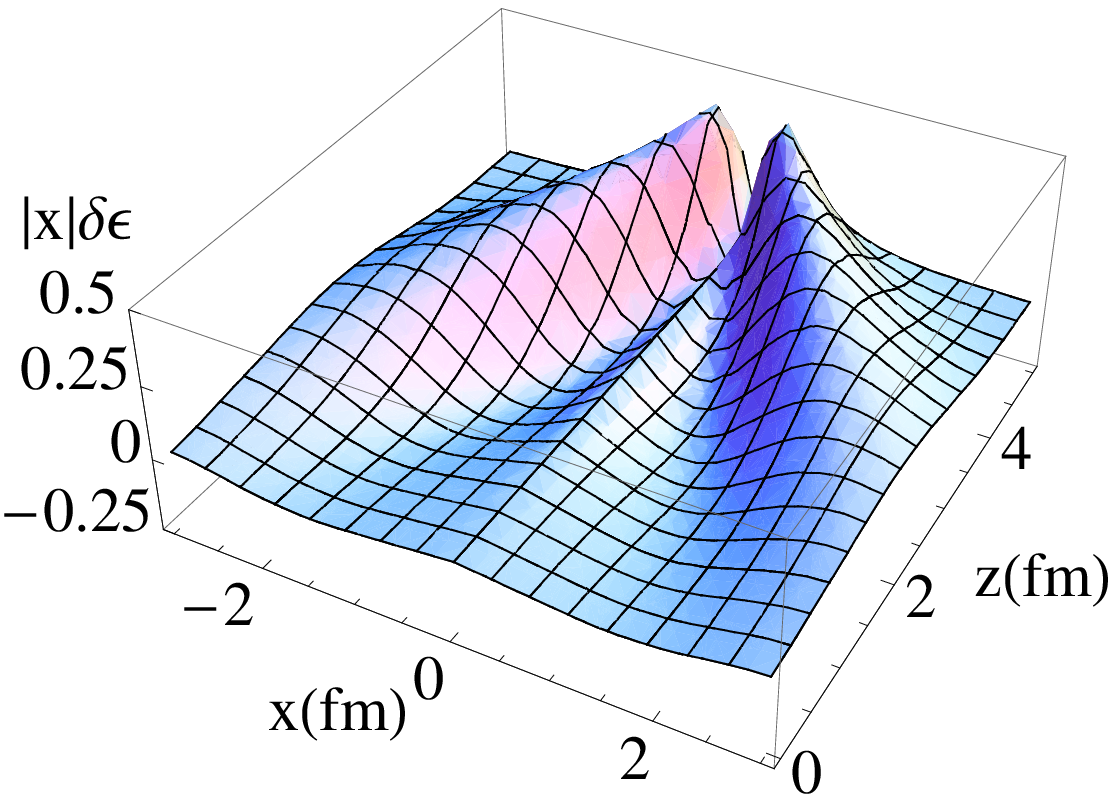}} 
\hspace{1.1cm}
\resizebox{1.1in}{1.1in}
          {\includegraphics[1in,0in][4in,3in]{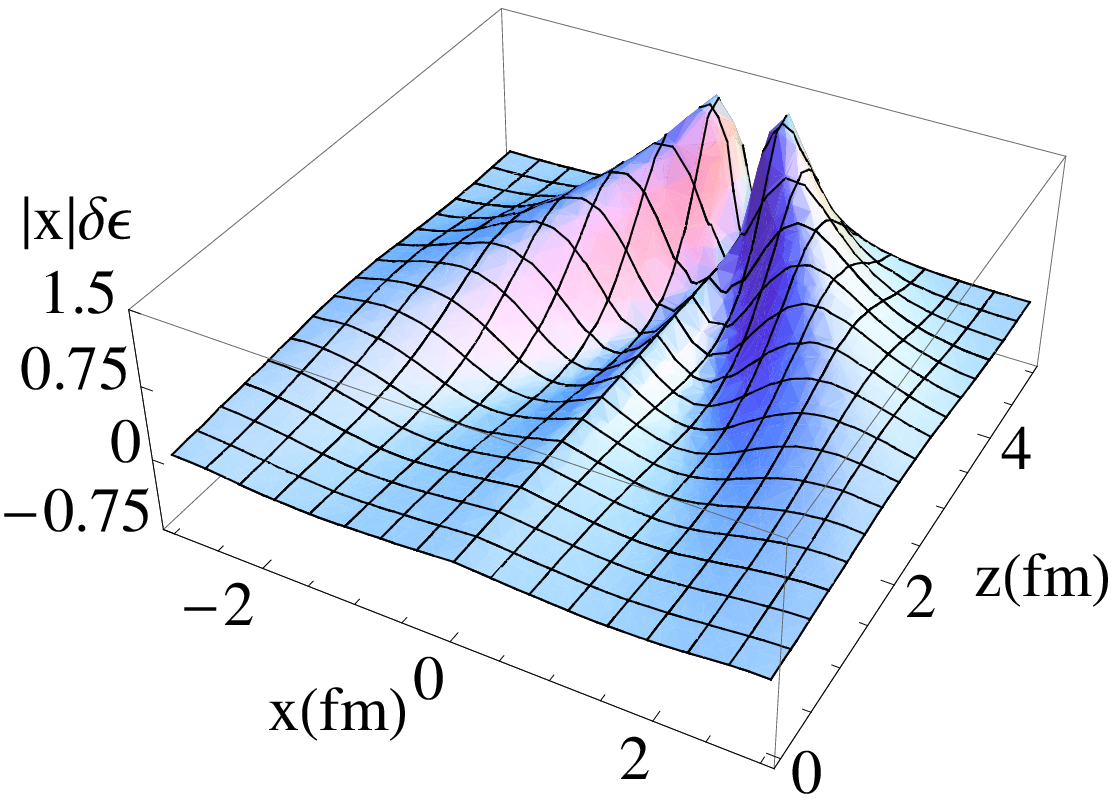}} 
\\
\mbx\vspace{0.2cm}\\
\hspace{0.4cm}
\resizebox{1.1in}{1.1in}
          {\includegraphics[1in,0in][4in,3in]{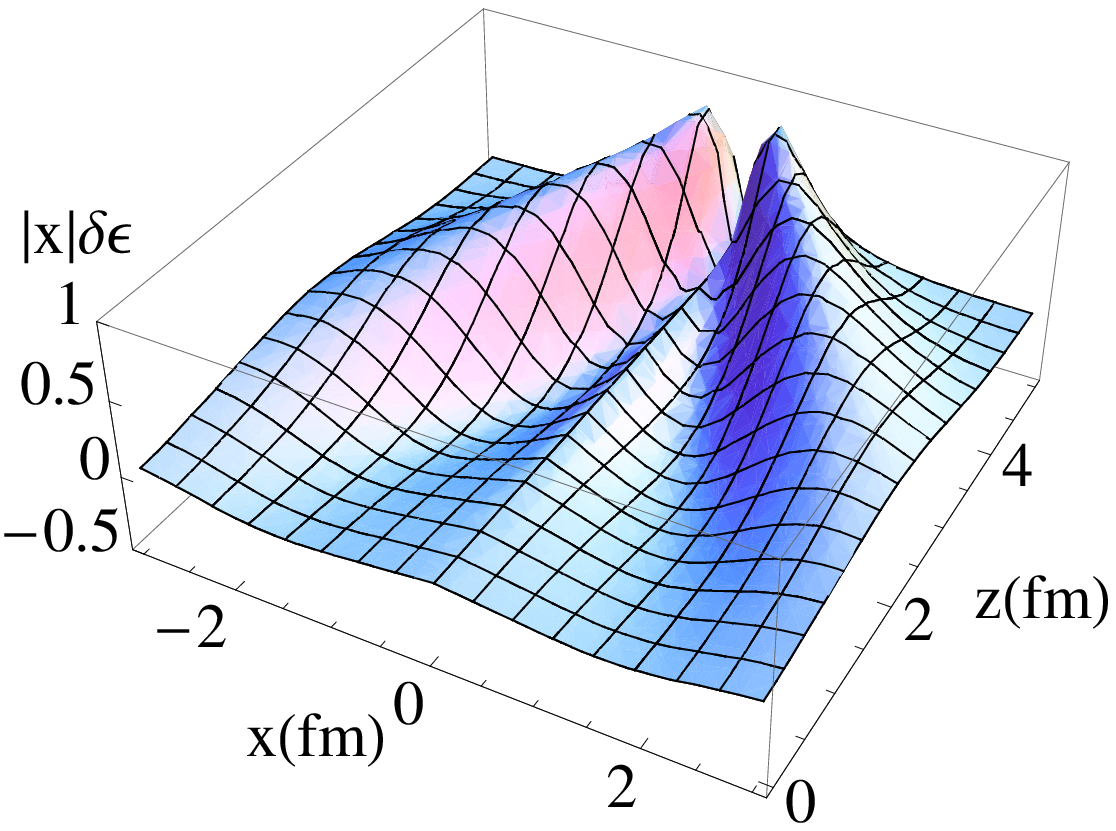}} 
\hspace{1.1cm}
\resizebox{1.1in}{1.1in}
          {\includegraphics[1in,0in][4in,3in]{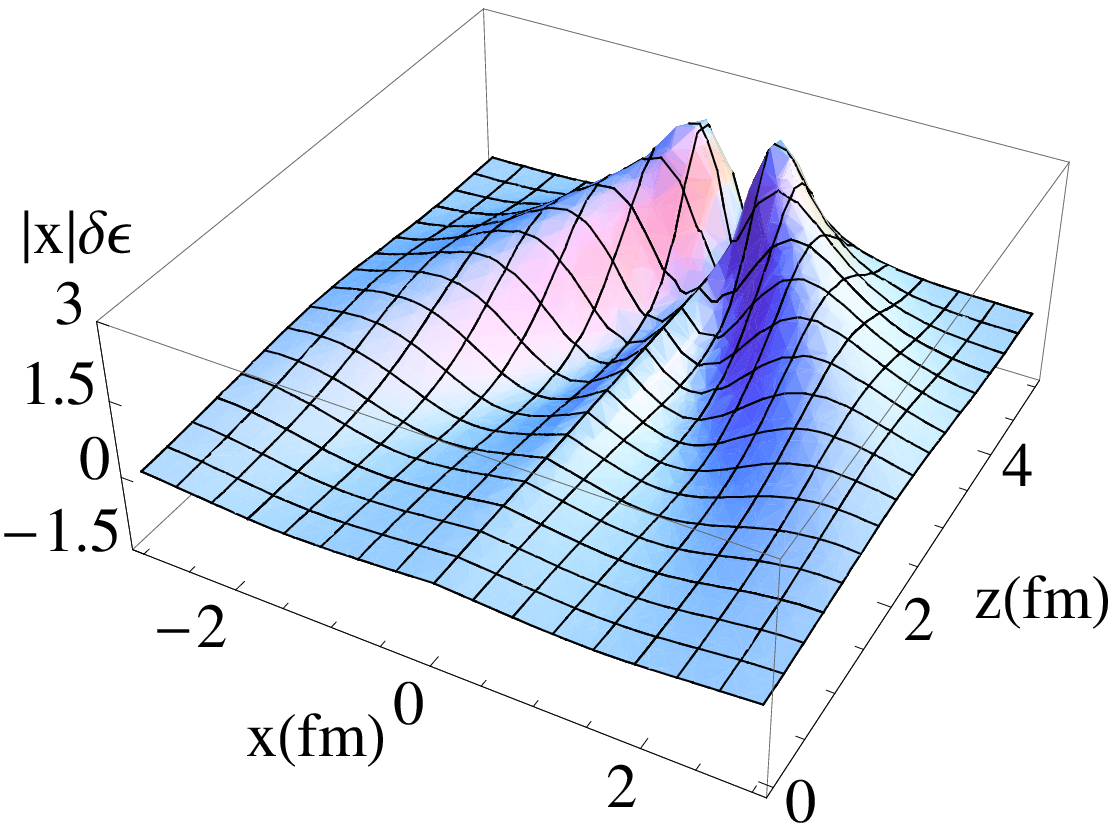}} 
\caption{(Color online) The linear fluid dynamical response to the energy 
deposited by a single parton (left) or by a parton-initiated shower (right), 
when the parton is a quark (top) or a gluon (bottom). Note the
different vertical scales.
} 
\label{fig4}
\end{figure}

In Fig.~\ref{fig4} we show the azimuthal projection of the energy 
density $|x|\kd \e$ at $t\eq5$fm/c after the parton is created, for a 
single non-radiating parton (left) and a parton-initiated jet shower 
(right). A gluon (bottom row) deposits more energy than a quark (top 
row), due to its larger color factor that enters both in the elastic 
energy loss and shower production rate. One immediately notes that, 
while the basic Mach cone structure is not changed, showering leads to 
an enhancement by a factor of 3 in the overall magnitude of the response. 
For quark jets our results are qualitatively similar to 
Ref.~\cite{Neufeld:2009ep}.

In this Letter, we have presented a consistent pQCD based calculation of 
the light-cone and transverse momentum ($\D q^-,p^2_T$) deposited by a 
jet in a medium, as a function of distance traversed. Assuming a short
thermalization time for the deposited energy we also computed the 
hydrodynamic response. The pQCD shower has the effect of a large part of 
the energy being deposited later in the history of the jet 
\cite{Neufeld:2009ep} which tends to enhance the Mach cone like 
structure formed. 

We thank B.~M\"{u}ller and R.~B.~Neufeld for helpful discussions. 
This work was supported by the U.S. Department of Energy under 
grant DE-FG02-01ER41190.  



\begin{thebibliography}{99}

\bibitem{RHIC_Whitepapers}
  I.~Arsene {\em et al.}, 
  Nucl.\ Phys.\ A {\bf 757}, 1 (2005);
  B.~B.~Back {\em et al.},
  {\em ibid.} {\bf 757}, 28 (2005);
  J.~Adams {\em et al.},
  {\em ibid.} {\bf 757}, 102 (2005);
  K.~Adcox {\em et al.},
  {\em ibid.} {\bf 757}, 184 (2005).

\bibitem{highpt}
K.~Adcox {\it et al.}, 
Phys.\ Rev.\ Lett.\  {\bf 88}, 022301 (2002);
C.~Adler {\it et al.}, 
Phys.\ Rev.\ Lett.\  {\bf 89}, 202301 (2002).

\bibitem{back_to_back} 
  A.~Adare {\it et al.},
  Phys.\ Rev.\ Lett.\  {\bf 98}, 232302 (2007);
  B.~I.~Abelev {\it et al.},  
  Phys.\ Rev.\ Lett.\  {\bf 102}, 052302 (2009).

\bibitem{Majumder:2007iu}
  A.~Majumder,
  J.\ Phys.\ G {\bf 34}, S377 (2007). 

\bibitem{hydro_mach}
  A.~K.~Chaudhuri and U.~Heinz,
  Phys.\ Rev.\ Lett.\  {\bf 97}, 062301 (2006);
  T.~Renk and J.~Ruppert,
  Phys.\ Lett.\  B {\bf 646}, 19 (2007);
  B.~Betz, {\it et al.}
  arXiv:0812.4401 [nucl-th].

\bibitem{string_mach}
  For an alternate approach that assumes that the hard parton couples 
  strongly to the medium see 
  P.~M.~Chesler and L.~G.~Yaffe,
  Phys.\ Rev.\ Lett.\  {\bf 99}, 152001 (2007);
  S.~S.~Gubser, S.~S.~Pufu and A.~Yarom,
  {\em ibid.} {\bf 100}, 012301 (2008).

\bibitem{Neufeld:2009ep}
  R.~B.~Neufeld and B.~M\"uller,
  Phys.\ Rev.\ Lett. {\bf 103}, 042301 (2009).
  
\bibitem{radiative} 
  U.~A.~Wiedemann,
  Nucl.\ Phys.\ B {\bf 588}, 303 (2000);
  C.~A.~Salgado and U.~A.~Wiedemann,
  Phys.\ Rev.\  D {\bf 68}, 014008 (2003).

\bibitem{Majumder:2004pt}
  A.~Majumder, E.~Wang and X.~N.~Wang,
  Phys.\ Rev.\ Lett.\  {\bf 99}, 152301 (2007).

\bibitem{AP}
V.~N.~Gribov and L.~N.~Lipatov,
Sov.\ J.\ Nucl.\ Phys.\  {\bf 15}, 438 (1972);
Yu.~L.~Dokshitzer,
Sov.\ Phys.\ JETP {\bf 46}, 641 (1977);
G.~Altarelli and G.~Parisi,
Nucl.\ Phys.\ B {\bf 126}, 298 (1977).
 
\bibitem{elastic}
  E.~Braaten and M.~H.~Thoma,
  Phys.\ Rev.\  D {\bf 44}, 2625 (1991).
  S.~Wicks, {\it et al.},
  Nucl.\ Phys.\  A {\bf 784}, 426 (2007);
  G.~Y.~Qin, {\it et al.},
  Phys.\ Rev.\ Lett.\  {\bf 100}, 072301 (2008).

\bibitem{Majumder:2008zg}
  A.~Majumder,
  arXiv:0810.4967 [nucl-th].

\bibitem{Majumder:2007hx}
  A.~Majumder and B.~M\"uller,
  Phys.\ Rev.\  C {\bf 77}, 054903 (2008).


\bibitem{Majumder:2009zu}
  A.~Majumder,
  arXiv:0901.4516 [nucl-th].


\bibitem{tobepublished}
  A. Majumder, to be published.


\bibitem{HT}
X.~F.~Guo and X.~N.~Wang,
Phys.\ Rev.\ Lett.\  {\bf 85}, 3591 (2000);
Nucl.\ Phys.\ A {\bf 696}, 788 (2001).

\bibitem{Majumder:2007ne}
  A.~Majumder, R.~J.~Fries and B.~M\"uller,
  Phys.\ Rev.\  C {\bf 77}, 065209 (2008).


\bibitem{linear_hydro}
  J.~Casalderrey-Solana, E.~V.~Shuryak and D.~Teaney,
  J.\ Phys.\ Conf.\ Ser.\  {\bf 27}, 22 (2005);
  R.~B.~Neufeld,
  Phys.\ Rev.\  C {\bf 79}, 054909 (2009).

\end{thebibliography}
\end{document}